\documentclass[groupedaddress, superscriptaddress, citesort]{revtex4}
\usepackage{amssymb, amsmath}
\usepackage[dvips]{graphicx}

\def\be{\begin{equation}}
\def\ee{\end{equation}}
\def\ba{\begin{array}}
\def\ea{\end{array}}

\begin{document}
\baselineskip=18pt

\title {The One-Way Information Deficit and Geometry for a Class of Two-qubit States}
\author{Yao-Kun Wang}
\affiliation{School of Mathematical Sciences,  Capital Normal
University,  Beijing 100048,  China}
\affiliation{College of Mathematics,  Tonghua Normal University,
 Tonghua 134001,  China}
\author{Teng Ma}
\affiliation{School of Mathematical Sciences,  Capital Normal
University,  Beijing 100048,  China}
\author{Bo Li}
\affiliation{Institute of Physics, Chinese Academy of Sciences, Beijing 100190, China}
\affiliation{Department of Mathematics and Computer,  Shangrao Normal University,
 Shangrao 334001,  China}
\author{Zhi-Xi Wang}
\affiliation{School of Mathematical Sciences,  Capital Normal
University,  Beijing 100048,  China}

\begin{abstract}
 The work deficit, as introduced by Jonathan Oppenheim \emph{et al }[Phys. Rev. Lett. \textbf{89},  180402 (2002)] is a good measure of the quantum correlations in a state and provides a new standpoint for understanding quantum non-locality. In this paper, we  analytically evaluate the one-way information deficit (OWID) for the Bell-diagonal states and a class of two-qubit states and further give the  geometry picture for OWID. The dynamic behavior of the OWID under decoherence channel is investigated and it is shown that the OWID of some classes of $X$ states is more robust against the decoherence than the entanglement.
\end{abstract}

\maketitle

\section{Introduction}

  The quantum entanglement of quantum states enables fascinating quantum information processing tasks such as super-dense coding\cite{CH}, teleportation\cite{CH2}, quantum cryptography\cite{AK},
remote-state preparation\cite{AK2} and so on.  However, quantum correlation other than entanglement has been attracting much attention recently \cite{bennett, zurek1,1modi,Auccaise,Giorgi,1Streltsov,modi2}. An outstanding and widely accepted quantity of them is quantum discord introduced by Oliver and Zurek and independently by Henderson and Vedral\cite{zurek1}. Quantum discord, a measure which quantify the difference between the  mutual information and maximum classical mutual information, is notorious difficult to calculate even for two qubit quantum system\cite{Ali,Li,chen,shi,Vinjanampathy}. The geometry of two qubit Bell-diagonal states was first introduced by Horodecki\cite{1Horodecki}. Recently, the geometry of  quantum discord for two qubit Bell-diagonal states and a class of five parameters X shape states reveals much difference than entanglement\cite{Li,langcaves}, and  some further investigation as geometric quantum discord appeared in\cite{yao}.

  There are also some  nonclassical correlation other than entanglement and quantum discord arising increasing interesting very recently. For example, the quantum deficit \cite{oppenheim,horodecki}, measurement-induced disturbance \cite{luo}, symmetric discord \cite{piani,wu}, relative entropy of discord and dissonance \cite{modi}, geometric discord \cite{luoandfu,dakic}, and continuous-variable discord \cite{adesso,giorda}, and so on\cite{modi2}. Among all of them, the work deficit\cite{oppenheim} is the first operational approach to quantify quantum correlations.  A similar physical interpretation of quantum discord appeared in\cite{zurek2} thereafter.  In the physical view, the results of \cite{oppenheim} shows that quantum deficit originates in questions using nonlocal operation to extract work from a correlated system coupled to a heat bath only in the case of pure states, and in the general case, the advantage is related to more general forms of quantum correlations. Jonathan Oppenheim \emph{et al } define the work deficit \cite{oppenheim}
\begin{eqnarray}
\Delta\equiv W_{t}-W_{l},
\end{eqnarray}
where $W_{t}$ is the information of the whole system and $W_{l}$ is the localizable information\cite{horodecki2}.
Similar to quantum discord,  quantum deficit is also equal to the difference of the mutual information and classical deficit \cite{oppenheim2}.  Recently, Alexander Streltsov \emph{et al }\cite{Streltsov0,chuan} give the definition of the one-way information deficit (OWID) by the relative entropy over all local von Neumann measurements on one subsystem, which reveals an important role of quantum correlations as a resource for the distribution of entanglement. The definition of the OWID by von Neumann measurements on one side is given by\cite{streltsov}
\begin{eqnarray}
\Delta^{\rightarrow}(\rho^{ab})=\min\limits_{\{\Pi_{k}\}}S(\sum\limits_{i}\Pi_{k}\rho^{ab}\Pi_{k})-S(\rho^{ab}).\label{definition}
\end{eqnarray}
From the definition we can find that the OWID and  quantum discord have similar minimum form but they are exactly different kinds of quantum correlation. A natural and interesting question is that can we obtain the analytical formula for some well known state such as Bell-diagonal states as like quantum discord? In this paper,
we endeavored to calculate the one-way  quantum deficit of  two qubit Bell-diagonal states and a class of four parameters X shape states. We first review some concepts for two qubit system. By using proper local unitary transformations,  we can write $\rho^{ab}$ as
\begin{eqnarray}
\rho^{ab}=\frac{1}{4}(I\otimes I+\textbf{r}\cdot\sigma\otimes I+I\otimes\textbf{s}\cdot\sigma+\sum_{i=1}^3c_i\sigma_i\otimes\sigma_i), \label{state1}
\end{eqnarray}
where \textbf{r} and \textbf{s} are Bloch vectors and $\{\sigma_i\}_{i=1}^3$ are
the standard Pauli matrices. When \textbf{r}=\textbf{s}=\textbf{0},
$\rho$ reduces to the two-qubit Bell-diagonal states.
Next,  we assume that the Bloch vectors are $z$ directional,  that is,  $\textbf{r}=(0, 0, r)$,  $\textbf{s}=(0, 0, s)$.
One can also change them to be $x$ or $y$ directional via
an appropriate local unitary transformation without losing its diagonal property of the correlation terms \cite{kim}.
This paper is organized as follows. In Section \ref{II}, we calculate the OWID defined by Eq.(\ref{definition}) for Bell-diagonal states and we can find that the OWID for Bell-diagonal states equals to its quantum discord. In Sec. \ref{III}, we depict the level surface of constant OWID in four different situations. In sec. \ref{IIII},  we discuss the dynamics of the OWID
and show that the OWID of a four parameters class of $X$ states  decay under decoherence channels.
A brief conclusion is given in sec. \ref{IIIII}.

\section{The OWID of Bell-diagonal States}\label{II}
For the two-qubit Bell-diagonal state
\begin{eqnarray}
\rho^{ab}=\frac{1}{4}(I\otimes I+\sum_{i=1}^3c_i\sigma_i\otimes\sigma_i).
\end{eqnarray}
The eigenvalues of $\rho^{ab}$ are given by
\begin{eqnarray}
\lambda_{1,2}=\frac{1}{4}(1-c_{1}\mp c_{2}\mp c_{3}),\quad
\lambda_{3,4}=\frac{1}{4}(1+c_{1}\mp c_{2}\pm c_{3}).\nonumber
\end{eqnarray}
The entropy of  $\rho^{ab}$ is
\begin{eqnarray}
S(\rho^{ab})
&=&-\sum_{i=1}^{4}\lambda_{i}\log \lambda_i\nonumber\\
&=&2-\frac{1-c_1-c_2-c_3}{4}\log (1-c_1-c_2-c_3)\nonumber\\
& &-\frac{1-c_1+c_2+c_3}{4}\log (1-c_1+c_2+c_3)\nonumber\\
& &-\frac{1+c_1-c_2+c_3}{4}\log (1+c_1-c_2+c_3)\nonumber\\
& &-\frac{1+c_1+c_2-c_3}{4}\log (1+c_1+c_2-c_3). \label{entropy1}
\end{eqnarray}
Next, we evaluate the OWID of the Bell-diagonal States. Let
\begin{eqnarray}
\{\Pi_{k}=|k\rangle\langle k|, k=0, 1\}
\end{eqnarray}
be the local measurement for the particle $b$ along the computational base ${|k\rangle}$;  then any von Neumann measurement for the particle $b$ can be written as
\begin{eqnarray}
\{B_{k}=V\Pi_{k}V^{\dag}: k=0, 1\}
\end{eqnarray}
for some unitary $V\in U(2)$. For any unitary $V$, we have
\begin{eqnarray}
V=tI+i\vec{y}\vec{\sigma}
\end{eqnarray}
with $t\in R$, $\vec{y}=(y_{1}, y_{2}, y_{3})\in R^{3}$, and $t^{2}+y_{1}^{2}+y_{2}^{2}+y_{3}^{2}=1. $
After the measurement ${B_{k}}$, the state $\rho^{ab}$ will be changed to the ensemble $\{{\rho_{k}, p_{k}}\}$ with
\begin{eqnarray}
\rho_{k}: =\frac{1}{p_{k}}(I\otimes B_{k})\rho(I\otimes B_{k})
\end{eqnarray}
and $p_{k}=tr(I\otimes B_{k})\rho(I\otimes B_{k})$. To evaluate $\rho_{k}$ and $p_{k}$, we write
\begin{eqnarray}
p_{k}\rho_{k}=(I\otimes B_{k})\rho(I\otimes B_{k})=\frac{1}{4}(I\otimes V)(I\otimes \Pi_{k})(I+\sum_{j=1}^{3} c_{j}\sigma_{j}\otimes (V^{\dag} \sigma_{j} V))(I\otimes \Pi_{k})(I\otimes V^{\dag}).
\end{eqnarray}

By the relations \cite{luo}
\begin{eqnarray}
V^{\dag}\sigma_{1}V=(t^{2}+y_{1}^{2}-y_{2}^{2}-y_{3}^{2})\sigma_{1}+2(ty_{3}+y_{1}y_{2})\sigma_{2}+2(-ty_{2}+y_{1}y_{3})\sigma_{3},\label{condition2} \\
V^{\dag}\sigma_{2}V=2(-ty_{3}+y_{1}y_{2})\sigma_{1}+(t^{2}+y_{2}^{2}-y_{1}^{2}-y_{3}^{2})\sigma_{2}+2(ty_{1}+y_{2}y_{3})\sigma_{3},\label{condition3} \\
V^{\dag}\sigma_{3}V=2(ty_{2}+y_{1}y_{3})\sigma_{1}+2(-ty_{1}+y_{2}y_{3})\sigma_{2}+(t^{2}+y_{3}^{2}-y_{1}^{2}-y_{2}^{2})\sigma_{3},\label{condition4}
\end{eqnarray}
and
\begin{eqnarray}
\Pi_{0}\sigma_{3}\Pi_{0}=\Pi_{0}, \Pi_{1}\sigma_{3}\Pi_{1}=-\Pi_{1}, \Pi_{j}\sigma_{k}\Pi_{j}=0, for j=0, 1, k=1, 2, \label{condition5}
 \end{eqnarray}
we obtain $p_{0}=p_{1}=\frac{1}{2}$ and
\begin{eqnarray}
\rho_{0}=\frac{1}{2}(I+c_{1}z_{1}\sigma_{1}+c_{2}z_{2}\sigma_{2}+c_{3}z_{3}\sigma_{3})\otimes(V\Pi_{0}V^{\dag}),\\
\rho_{1}=\frac{1}{2}(I-c_{1}z_{1}\sigma_{1}-c_{2}z_{2}\sigma_{2}-c_{3}z_{3}\sigma_{3})\otimes(V\Pi_{1}V^{\dag}),
\end{eqnarray}
where
\begin{eqnarray}
z_{1}=2(-ty_{2}+y_{1}y_{3}), \quad z_{2}=2(ty_{1}+y_{2}y_{3}), \quad z_{3}=t^{2}+y_{3}^{2}-y_{1}^{2}-y_{2}^{2}.\label{condition6}
\end{eqnarray}
Let $M=c_{1}z_{1}\sigma_{1}+c_{2}z_{2}\sigma_{2}+c_{3}z_{3}\sigma_{3}$, then
\begin{eqnarray}
\rho_{0}=\frac{1}{2}(I+M)\otimes(V\Pi_{0}V^{\dag}),\quad
\rho_{1}=\frac{1}{2}(I-M)\otimes(V\Pi_{1}V^{\dag}).
\end{eqnarray}
The eigenvalues of $\frac{1}{2}(I+M)$ and $\frac{1}{2}(I-M)$ are $\frac{1-\theta}{2}$, $\frac{1+\theta}{2}$ and $\frac{1+\theta}{2}$, $\frac{1-\theta}{2}$ respectively, where $\theta=\sqrt{|c_{1}z_{1}|^{2}+|c_{2}z_{2}|^{2}+|c_{3}z_{3}|^{2}}$.
 Since $\frac{1}{2}(I+M)$ commutes with $\frac{1}{2}(I-M)$,  there exists an orthogonal basis such that both $\frac{1}{2}(I+M)$ and  $\frac{1}{2}(I-M)$ are diagonal with respect to that basis \cite{nielsen}, that is there exits unitary $U\in U(2)$, and
\begin{eqnarray}
\frac{1}{2}(I+M)=U\left(\begin{array}{cc}\frac{1-\theta}{2} & 0 \\0 & \frac{1+\theta}{2} \end{array}\right)U^{\dag},\quad
\frac{1}{2}(I-M)=U\left(\begin{array}{cc}\frac{1+\theta}{2} & 0 \\0 & \frac{1-\theta}{2} \end{array}\right)U^{\dag}.
\end{eqnarray}

We evaluate the eigenvalues of $\sum\limits_{i}\Pi_{k}\rho^{ab}\Pi_{k}$ by $\sum\limits_{i}\Pi_{k}\rho^{ab}\Pi_{k}=p_{0}\rho_{0}+p_{1}\rho_{1}$,
\begin{eqnarray}
& &|\lambda E-\sum_{i}\Pi_{k}\rho^{ab}\Pi_{k}|\nonumber\\
&=&|\lambda E-(p_{0}\rho_{0}+p_{1}\rho_{1})|\nonumber\\
&=&\left|\lambda E-(\frac{1}{2}\rho_{0}+\frac{1}{2}\rho_{1})\right|\nonumber\\
&=&\left|\lambda E-\{[U\left(\begin{array}{cc}\frac{1-\theta}{4} & 0 \\0 & \frac{1+\theta}{4} \end{array}\right)U^{\dag}]\otimes(V\Pi_{0}V^{\dag})+[U\left(\begin{array}{cc}\frac{1+\theta}{4} & 0 \\0 & \frac{1-\theta}{4} \end{array}\right)U^{\dag}]\otimes(V\Pi_{1}V^{\dag})\}\right|\nonumber\\
&=&\left|\lambda E-\{(U\otimes V)[\left(\begin{array}{cc}\frac{1-\theta}{4} & 0 \\0 & \frac{1+\theta}{4} \end{array}\right)]\otimes\Pi_{0}](U^{\dag}\otimes V^{\dag})+\{(U\otimes V)[\left(\begin{array}{cc}\frac{1+\theta}{4} & 0 \\0 & \frac{1-\theta}{4} \end{array}\right)]\otimes\Pi_{1}](U^{\dag}\otimes V^{\dag})\}\right|\nonumber\\
&=&\left|\lambda E-(U\otimes V)[\left(\begin{array}{cc}\frac{1-\theta}{4} & 0 \\0 & \frac{1+\theta}{4} \end{array}\right)\otimes\Pi_{0}+\left(\begin{array}{cc}\frac{1+\theta}{4} & 0 \\0 & \frac{1-\theta}{4} \end{array}\right)\otimes\Pi_{1}](U^{\dag}\otimes V^{\dag})\right|\nonumber\\
&=&\left|(U\otimes V)\lambda E (U\otimes V)^{\dag}-(U\otimes V)\left(\begin{array}{cccc}\frac{1-\theta}{4} & 0& 0& 0 \\0 & \frac{1+\theta}{4}& 0& 0\\0& 0&\frac{1+\theta}{4}& 0\\0& 0& 0& \frac{1-\theta}{4} \end{array}\right)(U\otimes V)^{\dag}\right|\nonumber\\
&=&|(U\otimes V)|\cdot\left|\lambda E -\left(\begin{array}{cccc}\frac{1-\theta}{4} & 0& 0& 0 \\0 & \frac{1+\theta}{4}& 0& 0\\0& 0&\frac{1+\theta}{4}& 0\\0& 0& 0& \frac{1-\theta}{4} \end{array}\right)\right|\cdot|(U\otimes V)^{\dag}|\nonumber\\
&=&\left|\lambda E -\left(\begin{array}{cccc}\frac{1-\theta}{4} & 0& 0& 0 \\0 & \frac{1+\theta}{4}& 0& 0\\0& 0&\frac{1+\theta}{4}& 0\\0& 0& 0& \frac{1-\theta}{4} \end{array}\right)\right|. \label{eigenvalue1}
\end{eqnarray}
The eigenvalues of $\sum\limits_{i}\Pi_{k}\rho^{ab}\Pi_{k}$ are $\lambda_{5}=\frac{1-\theta}{4}$, $\lambda_{6}=\frac{1+\theta}{4}$, $\lambda_{7}=\frac{1+\theta}{4}$, $\lambda_{8}=\frac{1-\theta}{4}$, thus the entropy of $\sum\limits_{i}\Pi_{k}\rho^{ab}\Pi_{k}$ is given by
\begin{eqnarray}
S(\sum_{i}\Pi_{k}\rho^{ab}\Pi_{k})&=&-\sum_{i=5}^{8}\lambda_{i}\log \lambda_{i}\nonumber\\
&=&2-\frac{1-\theta}{2}\log (1-\theta)-\frac{1+\theta}{2}\log (1+\theta).
\end{eqnarray}
It can be directly verified that
\begin{eqnarray}
z_{1}^{2}+z_{2}^{2}+z_{3}^{2}=1.\label{condition7}
\end{eqnarray}
Let us put $c:=\max\{|c_{1}|, |c_{2}|, |c_{3}|\},$ then $\theta\leq\sqrt{|c|^{2}(|z_{1}|^{2}+|z_{2}|^{2}+|z_{3}|^{2})}=c$, and the equality can be readily attained by appropriate choice of $t,y_{j}$ \cite{luo}.
Therefore, we see that
\begin{eqnarray}
\sup\limits_{\{B_{k}\}}\theta=\sup\limits_{\{V\}}\theta=c.
\end{eqnarray}
The range of values allowed for $\theta$ is $[0, c]$,  and the derivative of $S(\sum\limits_{i}\Pi_{k}\rho^{ab}\Pi_{k})$ is $\frac{1}{2}[\log (1-\theta)-\log (1+\theta)]$, and it can be verified that it is a monotonic decreasing function in the interval $[0, c]$,  and the minimal value of $S(\sum\limits_{i}\Pi_{k}\rho^{ab}\Pi_{k})$ can be attained at point $c$,
\begin{eqnarray}
\min\limits_{\{\Pi_{k}\}}S(\sum_{i}\Pi_{k}\rho^{ab}\Pi_{k})=2-\frac{1-c}{2}\log (1-c)-\frac{1+c}{2}\log (1+c). \label{min1}
\end{eqnarray}
By Eqs.(\ref{entropy1}), (\ref{min1}), and the OWID of $\rho^{ab}$ is given by
\begin{eqnarray}
 \Delta^{\rightarrow}(\rho^{ab})&=&\min\limits_{\{\Pi_{k}\}}S(\sum\limits_{i}\Pi_{k}\rho^{ab}\Pi_{k})-S(\rho^{ab})\nonumber\\
&=&\frac{1}{4}[(1-c_1-c_2-c_3)\log (1-c_1-c_2-c_3)\nonumber\\
& &+(1-c_1+c_2+c_3)\log (1-c_1+c_2+c_3)\nonumber\\
& &+(1+c_1-c_2+c_3)\log (1+c_1-c_2+c_3)\nonumber\\
& &+(1+c_1+c_2-c_3)\log (1+c_1+c_2-c_3)]\nonumber\\
& &-\frac{1-c}{2}\log (1-c)-\frac{1+c}{2}\log (1+c).
\end{eqnarray}

It should be mentioned that the above result about the OWID of the Bell-diagonal states coincides with its quantum discord obtained in \cite{luo}. The geometry picture  of the OWID given in Figure 2 of \cite{langcaves}.

\section{The OWID of a class of $X$ states and its geometrical depiction}\label{III}
Although it is well-known that quantum discord in \cite{luo} can be extended to the full 5-parameter family \cite{Li} and the full 7-parameter  family of X-states \cite{Ali},  all these are special cases and actually even further to all qubit states (15 parameters) and even more \cite{Vinjanampathy}, the serious difficulty is that one is not even able to obtain the value of the OWID of the full 5-parameter family of X-states. Here we will evaluate the full 4-parameter family of X-states with additional assumptions.

We consider the following 4-parameter quantum system,
\begin{widetext}
\begin{eqnarray}
\rho^{ab} = \frac{1}{4} \left(
\begin{array}{cccc}
1+s+c_3
& 0 & 0 & c_1 -c_2 \\
0 & 1-s-c_3 & c_1+c_2 & 0 \\
0 & c_1 +c_2 & 1+s-c_3
& 0 \\
c_1 -c_2 & 0 & 0 & 1-s+c_3
\end{array}
\right), \label{matrix2}
\end{eqnarray}
\end{widetext}
we will only consider the following further simplified family of the Eq.(\ref{matrix2}), where
\begin{eqnarray}
|c_{1}|<|c_{2}|<|c_{3}|, \quad 0<|s|<1-|c_{3}|. \label{condition1}
\end{eqnarray}

The concurrence of the state in Eqs.(\ref{matrix2}), (\ref{condition1}) can be calculated in terms of the eigenvalues of $\rho\widetilde{\rho}$,  where $\widetilde{\rho}=\sigma_y\otimes \sigma_y\rho^*\sigma_y\otimes \sigma_y$.
The eigenvalues of $\rho\widetilde{\rho}$ are
\begin{eqnarray}
\lambda_{9}&=&\frac{1}{16}(c_1-c_2-\sqrt{(1+c_3)^2-s^2})^2\nonumber\\
&=&\frac{1}{16}(c_1-c_2-\sqrt{(1+s+c_3)(1-s+c_3)})^2,\nonumber
\end{eqnarray}
\begin{eqnarray}
\lambda_{10}&=&\frac{1}{16}(c_1-c_2+\sqrt{(1+c_3)^2-s^2})^2\nonumber\\
&=&\frac{1}{16}(c_1-c_2+\sqrt{(1+s+c_3)(1-s+c_3)})^2,\nonumber
\end{eqnarray}
\begin{eqnarray}
\lambda_{11}&=&\frac{1}{16}(c_1+c_2-\sqrt{(1-c_3)^2-s^2})^2\nonumber\\
&=&\frac{1}{16}(c_1+c_2-\sqrt{(1-s-c_3)(1+s-c_3)})^2,\nonumber
\end{eqnarray}
\begin{eqnarray}
\lambda_{12}&=&\frac{1}{16}(c_1+c_2+\sqrt{(1-c_3)^2-s^2})^2\nonumber\\
&=&\frac{1}{16}(c_1+c_2+\sqrt{(1-s-c_3)(1+s-c_3)})^2.\nonumber
\end{eqnarray}
The concurrence of the state in Eqs.(\ref{matrix2}), (\ref{condition1}) is given by
\begin{widetext}
\begin{eqnarray}
C(\rho^{ab})=\max\{2\max\{\sqrt{\lambda_{9}}, \sqrt{\lambda_{10}}, \sqrt{\lambda_{11}}, \sqrt{\lambda_{12}}\}
-\sqrt{\lambda_{9}}-\sqrt{\lambda_{10}}-\sqrt{\lambda_{11}}-\sqrt{\lambda_{12}}, 0 \}.
\label{twoqubitconcurrence}
\end{eqnarray}
\end{widetext}

We will only consider the OWID of the state in Eqs.(\ref{matrix2}), (\ref{condition1}). Our computation procedure of the OWID is similar to the Bell-diagonal state case. The eigenvalues of the state in Eq.(\ref{matrix2}), (\ref{condition1}) are given by
\begin{eqnarray}
\lambda_{13, 14}=\frac{1}{4}[1-c_3\pm\sqrt{s^2+(c_1+c_2)^2} ], \quad\lambda_{15, 16}=\frac{1}{4}[1+c_3\pm\sqrt{s^2+(c_1-c_2)^2} ].\nonumber
\end{eqnarray}
The entropy is given by
\begin{eqnarray}
S(\rho^{ab})&=&-\sum\limits_{i=13}^{16}\lambda_{i}\log \lambda_{i}\nonumber\\
&=&2-\frac{1}{4}[(1-c_{3}+\sqrt{s^{2}+(c_{1}+c_{2})^{2}})\log (1-c_{3}+\sqrt{s^{2}+(c_{1}+c_{2})^{2}})\nonumber\\
&&+(1-c_{3}-\sqrt{s^{2}+(c_{1}+c_{2})^{2}})\log (1-c_{3}-\sqrt{s^{2}+(c_{1}+c_{2})^{2}})\nonumber\\
&&+(1+c_{3}+\sqrt{s^{2}+(c_{1}+c_{2})^{2}})\log (1+c_{3}+\sqrt{s^{2}+(c_{1}+c_{2})^{2}})\nonumber\\
&&+(1+c_{3}-\sqrt{s^{2}+(c_{1}+c_{2})^{2}})\log (1+c_{3}-\sqrt{s^{2}+(c_{1}+c_{2})^{2}})]. \label{entropy2}
\end{eqnarray}
We need to evaluate the eigenvalues of $\sum\limits_{i}\Pi_{k}\rho^{ab}\Pi_{k}$ by $\sum\limits_{i}\Pi_{k}\rho^{ab}\Pi_{k}=p_{0}\rho_{0}+p_{1}\rho_{1}$. For this purpose, we write
\begin{eqnarray}
p_{k}\rho_{k}&=&(I\otimes B_{k})\rho(I\otimes B_{k})\nonumber\\
&=&\frac{1}{4}(I\otimes V)(I\otimes \Pi_{k})(I\otimes V^{\dag})(I\otimes I+I\otimes s\sigma_{3}+\sum_{i=1}^{3} c_{i}\sigma_{i}\otimes\sigma_{i}))(I\otimes V)(I\otimes \Pi_{k})(I\otimes V^{\dag})\nonumber\\
&=&\frac{1}{4}I\otimes V\Pi_{k}V^{\dag}+\frac{s}{4}I\otimes V\Pi_{k}V^{\dag}\sigma_{3}V\Pi_{k}V^{\dag}+\sum\limits_{i=1}^{3}\frac{c_{i}}{4}\sigma_{i}\otimes V\Pi_{k}V^{\dag}\sigma_{i}V\Pi_{k}V^{\dag}.
\end{eqnarray}
By the Eqs.(\ref{condition2}), (\ref{condition3}), (\ref{condition4}), (\ref{condition5}), (\ref{condition6}), we obtain
\begin{eqnarray}
p_{0}\rho_{0}=\frac{1}{4}(I+sz_{3}I+c_{1}z_{1}\sigma_{1}+c_{2}z_{2}\sigma_{2}+c_{3}z_{3}\sigma_{3})\otimes V\Pi_{0}V^{\dag}, \\
p_{1}\rho_{1}=\frac{1}{4}(I-sz_{3}I-c_{1}z_{1}\sigma_{1}-c_{2}z_{2}\sigma_{2}-c_{3}z_{3}\sigma_{3})\otimes V\Pi_{1}V^{\dag}.
\end{eqnarray}
Let $\overline{M}=sz_{3}I+c_{1}z_{1}\sigma_{1}+c_{2}z_{2}\sigma_{2}+c_{3}z_{3}\sigma_{3}$, and
\begin{eqnarray}
p_{0}\rho_{0}=\frac{1}{4}(I+\overline{M})\otimes V\Pi_{0}V^{\dag}, \quad
p_{1}\rho_{1}=\frac{1}{4}(I-\overline{M})\otimes V\Pi_{1}V^{\dag}.
\end{eqnarray}
Similar to Eq.(\ref{eigenvalue1}), the eigenvalues of $\frac{1}{4}(I+\overline{M})$ and $\frac{1}{4}(I-\overline{M})$ are $\lambda_{17}=\frac{1}{4}(1+\phi-\theta), \lambda_{18}=\frac{1}{4}(1+\phi+\theta)$ and $\lambda_{19}=\frac{1}{4}(1-\phi-\theta), \lambda_{20}=\frac{1}{4}(1-\phi+\theta), $ where
\begin{eqnarray}
\phi=sz_{3},   \theta=\sqrt{|c_{1}z_{1}|^{2}+|c_{2}z_{2}|^{2}+|c_{3}z_{3}|^{2}}.\label{condition9}
\end{eqnarray}
The entropy of $\sum\limits_{i}\Pi_{k}\rho^{ab}\Pi_{k}$
is
\begin{eqnarray}
S(\sum\limits_{i}\Pi_{k}\rho^{ab}\Pi_{k})&=&f(\phi,\theta )=-\sum\limits_{i=17}^{20}\lambda_{i}\log \lambda_{i}\nonumber\\
&=&2-\frac{1}{4}[(1+\phi-\theta)\log (1+\phi-\theta)+(1+\phi+\theta)\log (1+\phi+\theta)\nonumber\\
& &+(1-\phi-\theta)\log (1-\phi-\theta)+(1-\phi+\theta)\log (1-\phi+\theta)].
\end{eqnarray}
 By use of the domain scope of logarithmic function in $f(\phi, \theta)$ and Eq.(\ref{condition1}), we obtain the range of $\theta$ and $\phi$:
\begin{eqnarray}
0\leq|c_{1}|\leq\theta\leq|c_{3}|\leq1,\ \   -1<\phi<1.
\end{eqnarray}
We can verify that $f(-\phi, \theta)=f(\phi, \theta)$, the graph of $f(\phi, \theta)$ is symmetrical with respect to the $\theta$-axis; $\frac{\partial f}{\partial \theta}=-\frac{1}{4}\log [\frac{(1+\theta)^{2}-\phi^{2}}{(1-\theta)^{2}-\phi^{2}}]<0,\  0<\theta<1, $ $f(\phi, \theta)$ is a monotonic decreasing function; $\frac{\partial f}{\partial\phi}=-\frac{1}{4}\log [\frac{(1+\phi)^{2}-\theta^{2}}{(1-\phi)^{2}-\theta^{2}}]<0, \,\, 0<\phi<1, $
$f(\phi, \theta)$ is a  monotonic decreasing function.
When $\theta=|c_{3}|$, by Eqs.(\ref{condition7}), (\ref{condition1}), (\ref{condition9}), we can obtain
\begin{eqnarray}
\phi=|s|.
\end{eqnarray}
By Eq.(\ref{condition1}), the projection of $f(\phi, \theta)$ on the plane $\phi o\theta$  is  a symmetrical rectangle with respect to the $\theta$-axis, and by use of the monotonicity of $f(\phi, \theta)$ in the positive direction of $\theta$ and $\phi$,  $f(\phi, \theta)$ can obtain the minimum at the point $(|s|, |c_{3}|)$, the minimum of $f(\phi, \theta)$ is given by
\begin{eqnarray}
\min S(\sum\limits_{i}\Pi_{k}\rho^{ab}\Pi_{k})
&=&2-\frac{1}{4}[(1+s-c_{3})\log (1+s-c_{3})+(1+s+c_{3})\log (1+s+c_{3})\nonumber\\
& &+(1-s-c_{3})\log (1-s-c_{3})+(1-s+c_{3})\log (1-s+c_{3})].\label{min2}
\end{eqnarray}
By Eq.(\ref{entropy2}), (\ref{min2}), the OWID of the state in Eq.(\ref{matrix2}), (\ref{condition1}) is given by
\begin{eqnarray}
\Delta^{\rightarrow}(\rho^{ab})&=&\min\limits_{\{\Pi_{k}\}}S(\sum\limits_{i}\Pi_{k}\rho^{ab}\Pi_{k})-S(\rho^{ab})\nonumber\\
&=&\frac{1}{4}[(1-c_{3}+\sqrt{s^{2}+(c_{1}+c_{2})^{2}})\log (1-c_{3}+\sqrt{s^{2}+(c_{1}+c_{2})^{2}})\nonumber\\
& &+(1-c_{3}-\sqrt{s^{2}+(c_{1}+c_{2})^{2}})\log (1-c_{3}-\sqrt{s^{2}+(c_{1}+c_{2})^{2}})\nonumber\\
& &+(1+c_{3}+\sqrt{s^{2}+(c_{1}+c_{2})^{2}})\log (1+c_{3}+\sqrt{s^{2}+(c_{1}+c_{2})^{2}})\nonumber\\
& &+(1+c_{3}-\sqrt{s^{2}+(c_{1}+c_{2})^{2}})\log (1+c_{3}-\sqrt{s^{2}+(c_{1}+c_{2})^{2}})]\nonumber\\
& &-\frac{1}{4}[(1+s-c_{3})\log (1+s-c_{3})+(1+s+c_{3})\log (1+s+c_{3})\nonumber\\
& &+(1-s-c_{3})\log (1-s-c_{3})+(1-s+c_{3})\log (1-s+c_{3})].
\end{eqnarray}

In Fig.1 we plot the level surface of the OWID when (a) $s=0.3$,  $\Delta^{\rightarrow}(\rho^{ab})=0.03$;
(b) $s=0.5$,  $\Delta^{\rightarrow}(\rho^{ab})=0.03$; (c) $s=0.3$,  $\Delta^{\rightarrow}(\rho^{ab})=0.15$;
(d) $s=0.5$,  $\Delta^{\rightarrow}(\rho^{ab})=0.15$.
From Fig.1  one can see that the level surface of the OWID has a great change from the Bell diagonal states studied in Figure 2 of \cite{langcaves}. The surface shrinks with the effect $s$ and the shrinking rate becomes larger with the increasing $|s|$. What is more,  for larger deficit and small $s$ (see Fig.(c)),  the figure is similar to the ones in case of the Bell diagonal states. But for larger $s$ (see Fig.(d)),  the figure is moved up again and changes dramatically also.
\begin{figure}[h]
\raisebox{17em}{(a)}\includegraphics[width=6.25cm]{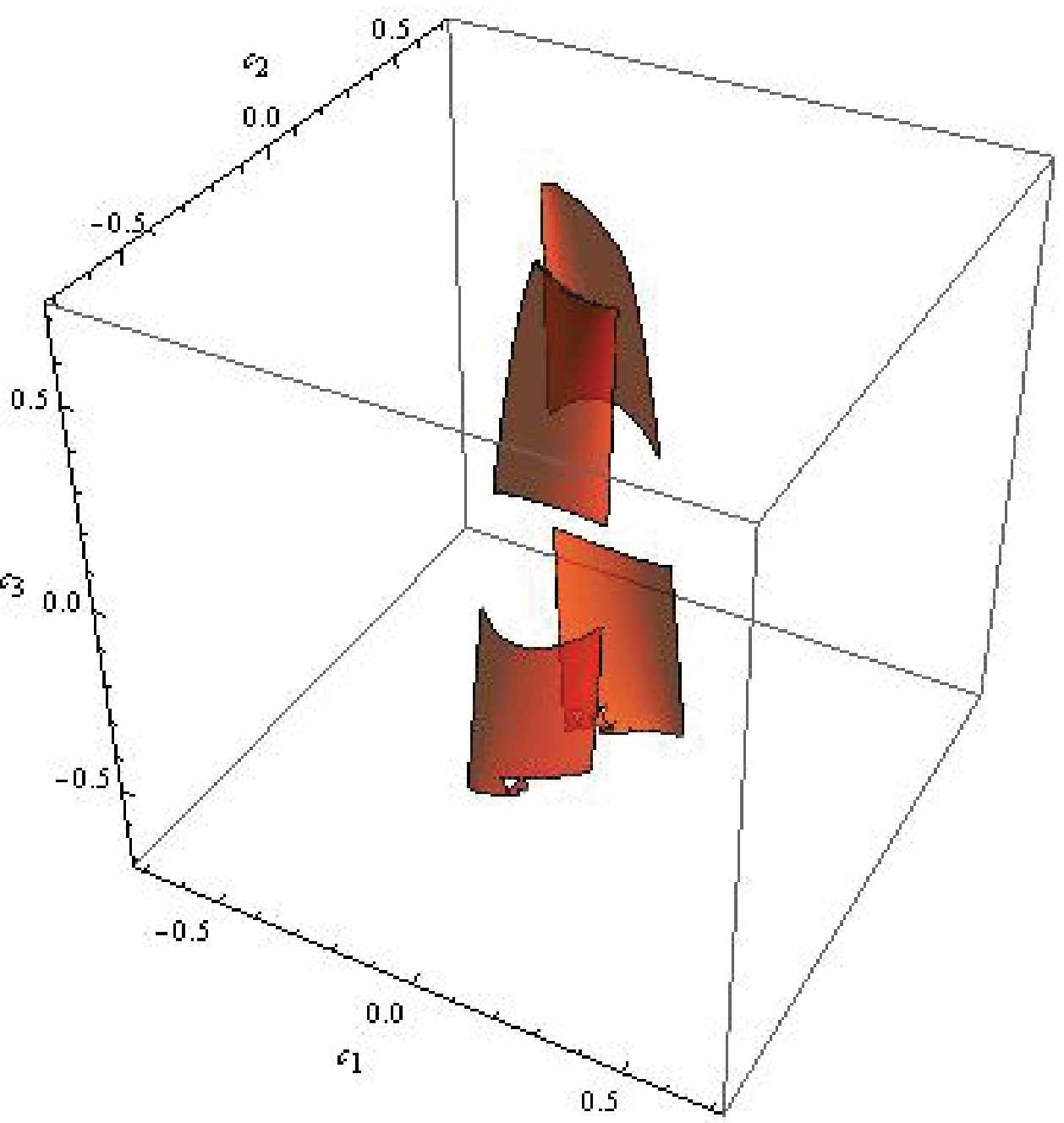}
\qquad
\raisebox{17em}{(b)}\includegraphics[width=6.25cm]{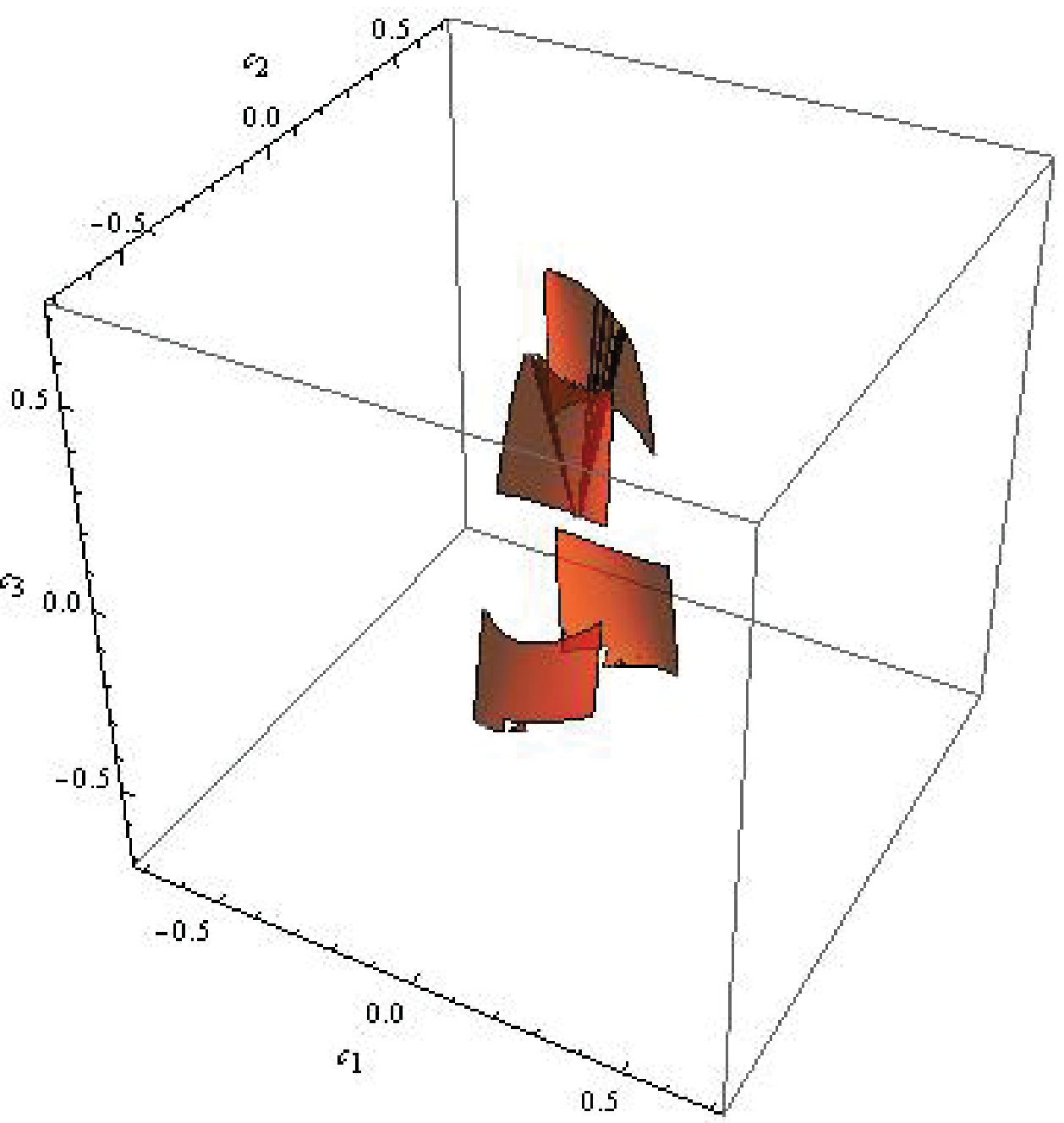}
\label{Fig:1}
\end{figure}
\begin{figure}[h]
\begin{center}
\raisebox{17em}{(c)}\includegraphics[width=6.25cm]{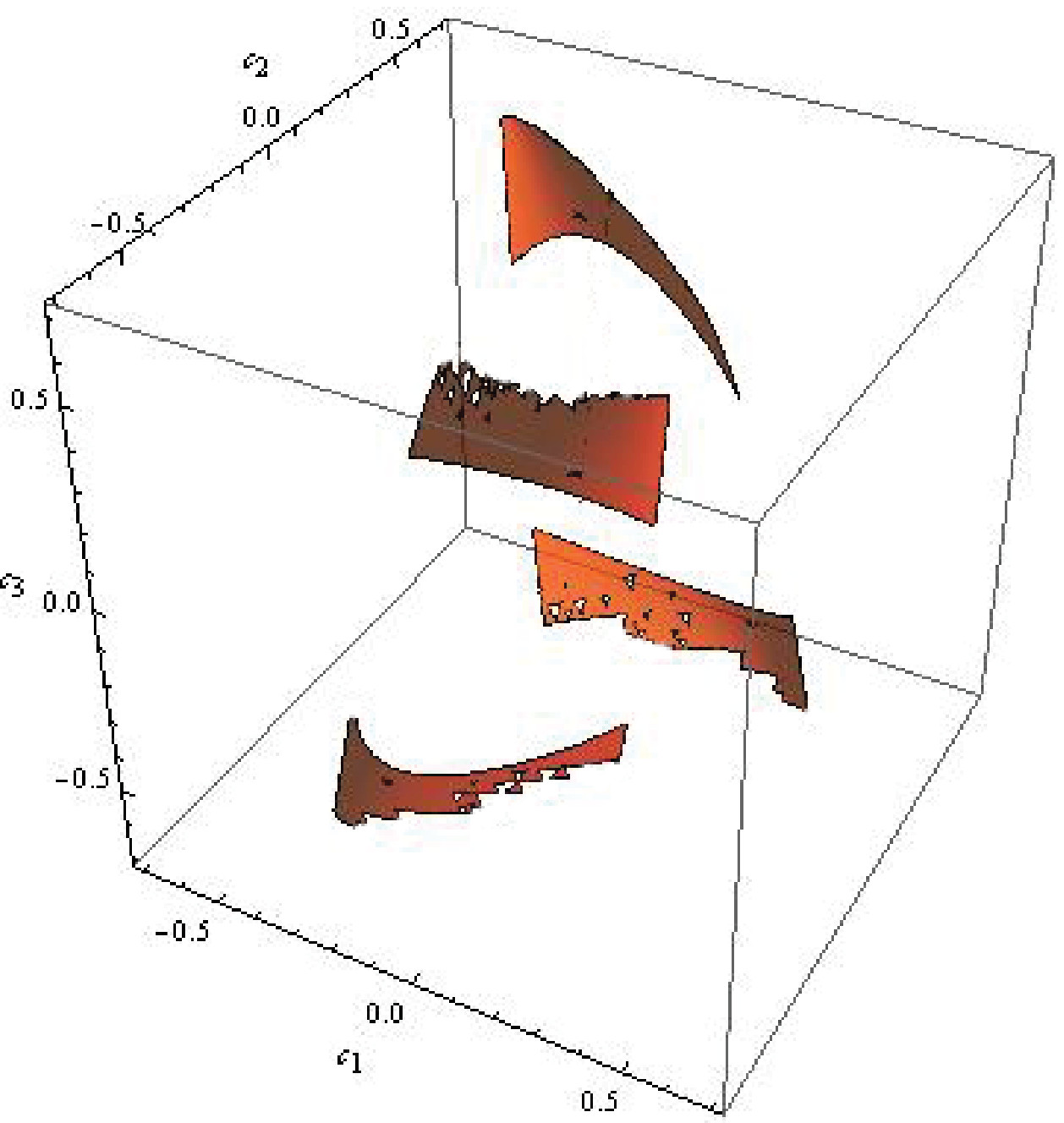}
\qquad
\raisebox{17em}{(d)}\includegraphics[width=6.25cm]{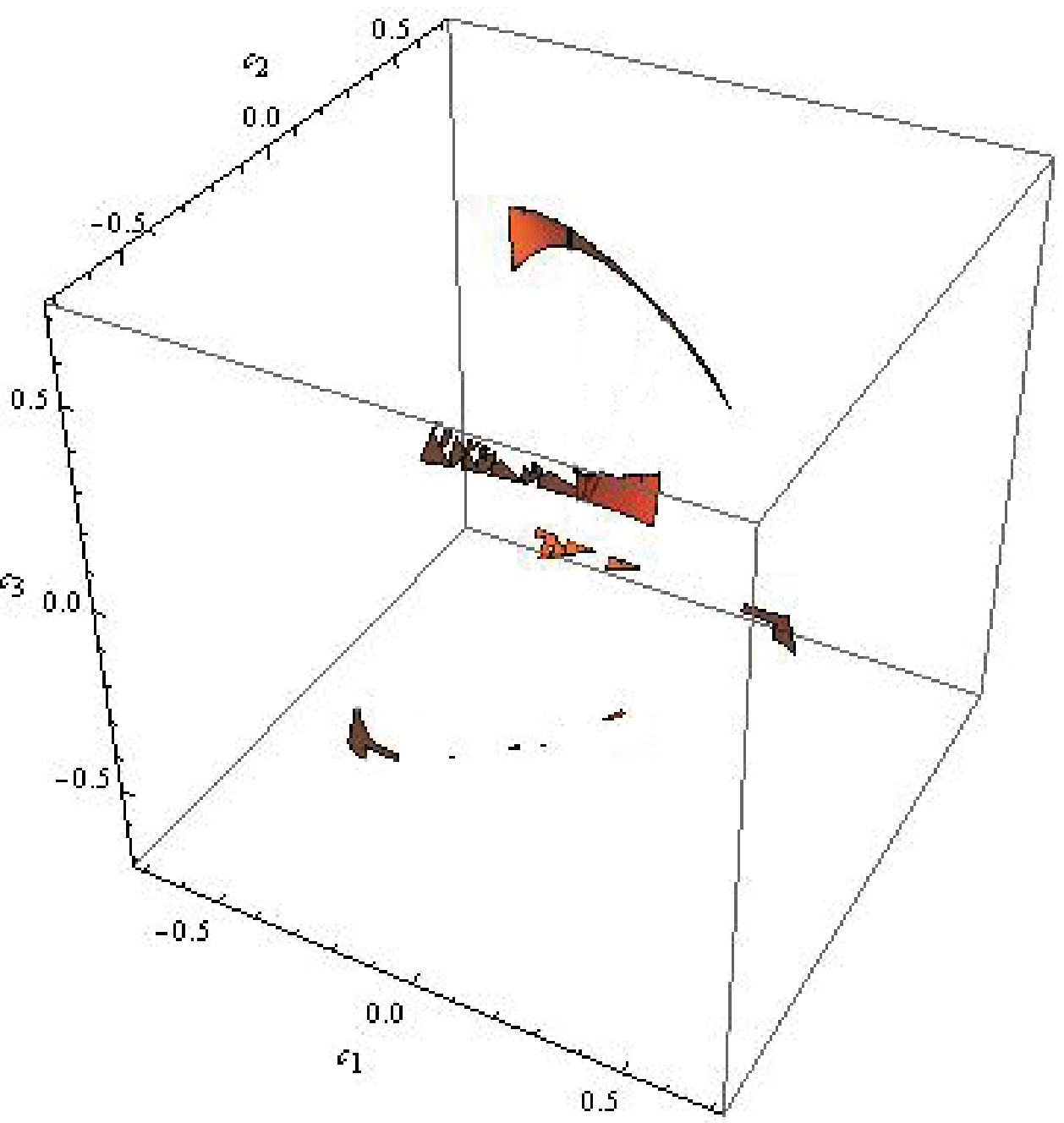}
\caption{(Color online) Surfaces of constant OWID: (a) $s=0.3$,  $\Delta^{\rightarrow}(\rho^{ab})=0.03$;
(b) $s=0.5$,  $\Delta^{\rightarrow}(\rho^{ab})=0.03$; (c) $s=0.3$,  $\Delta^{\rightarrow}(\rho^{ab})=0.15$;
(d) $s=0.5$,  $\Delta^{\rightarrow}(\rho^{ab})=0.15$.}
\end{center}
\end{figure}
\section{\bf Dynamics of OWID under local nondissipative channels}\label{IIII}
In the following we consider that the state in Eqs.(\ref{matrix2}), (\ref{condition1}) undergoes the phase flip channel \cite{Maziero},  with the Kraus operators
$\Gamma_0^{(A)}=$ diag$(\sqrt{1-p/2}, \sqrt{1-p/2})\otimes I$,  $\Gamma_1^{(A)}=$ diag$(\sqrt{p/2}, -\sqrt{p/2})\otimes I$,
$\Gamma_0^{(B)}= I \otimes$ diag$(\sqrt{1-p/2}, \sqrt{1-p/2}) $,  $\Gamma_1^{(B)}= I \otimes$ diag$(\sqrt{p/2}, -\sqrt{p/2}) $,  where $p=1-\exp(-\gamma t)$,  $\gamma$ is
the phase damping rate \cite{Maziero, yu}. Let $\varepsilon(\cdot)$ represent the operator of decoherence. Then under the phase flip channel  we have
\begin{eqnarray}
\varepsilon(\rho)&=& \frac{1}{4}(I\otimes I+I\otimes s \sigma_3+(1-p)^2c_1\sigma_1\otimes\sigma_1\nonumber\\
    &&+(1-p)^2c_2\sigma_2\otimes\sigma_2+c_3\sigma_3\otimes\sigma_3).
\end{eqnarray}

As $\varepsilon(\rho)$ satisfies conditions in Eqs.(\ref{matrix2}), (\ref{condition1}), and the OWID of the $\rho^{ab}$ under the phase flip channel is given by
\begin{eqnarray}
\Delta^{\rightarrow}(\varepsilon(\rho^{ab}))
&=&\frac{1}{4}[(1-c_{3}+\sqrt{s^{2}+(1-p)^{4}(c_{1}+c_{2})^{2}})\log (1-c_{3}+\sqrt{s^{2}+(1-p)^{4}(c_{1}+c_{2})^{2}})\nonumber\\
& &+(1-c_{3}-\sqrt{s^{2}+(1-p)^{4}(c_{1}+c_{2})^{2}})\log (1-c_{3}-\sqrt{s^{2}+(1-p)^{4}(c_{1}+c_{2})^{2}})\nonumber\\
& &+(1+c_{3}+\sqrt{s^{2}+(1-p)^{4}(c_{1}+c_{2})^{2}})\log (1+c_{3}+\sqrt{s^{2}+(1-p)^{4}(c_{1}+c_{2})^{2}})\nonumber\\
& &+(1+c_{3}-\sqrt{s^{2}+(1-p)^{4}(c_{1}+c_{2})^{2}})\log (1+c_{3}-\sqrt{s^{2}+(1-p)^{4}(c_{1}+c_{2})^{2}})]\nonumber\\
& &-\frac{1}{4}[(1+s-c_{3})\log (1+s-c_{3})+(1+s+c_{3})\log (1+s+c_{3})\nonumber\\
& &+(1-s-c_{3})\log (1-s-c_{3})+(1-s+c_{3})\log (1-s+c_{3})].
\end{eqnarray}

As an example, for $s=0.3, c_{1}=0.3, c_{2}=-0.4, c_{3}=0.56$, the dynamic behavior of correlation of the state under the phase flip channel is depicted in Fig.2. We find that the concurrence $C$ is greater than the OWID for $0\leq P\leq0.237211$ and a sudden death of entanglement appears at $p=0.321904, $ here one sees that the concurrence become zero after the transition. Therefore for these states the entanglement is weaker against the decoherence than the OWID.

\begin{figure}[h]
\scalebox{2.0}{\includegraphics[width=3.25cm]{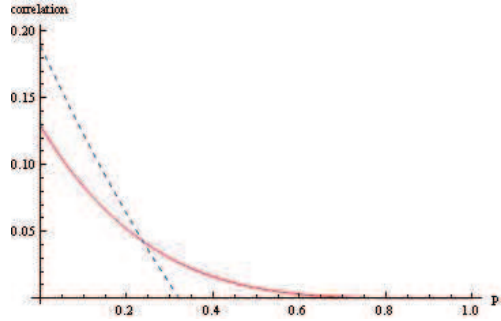}}
\caption{(Color online) Concurrence(blue dashed line) and OWID(red solid line) under phase flip channel for
$s=0.3$,  $c_1=0.3, $ $c_2=-0.4$ and $c_3=0.56$.}
\label{transition}
\end{figure}
\section{\bf summary}\label{IIIII}
We have studied the OWID for a class of $X$ states.
The level surfaces of the OWID have been depicted. For $r=s=0$ our results reduce to
the ones for Bell-diagonal states. For nonzero $s$,  it has been shown that the level surfaces of the OWID may have quite different geometry. The OWID become smaller in certain time interval for some initial states, and for some states the OWID is stronger against the decoherence than the entanglement.

\bigskip
\noindent {\bf Acknowledgments}  This work was supported by the NSFC 11101017, KZ201210028032 and PHR201007107.

\end{document}